\documentclass[dvips]{article}
\usepackage[dvips]{color}
\usepackage{graphicx} 
\usepackage{subeqn}

\renewcommand{\thesection}{\arabic{section}}

\renewcommand{\appendix}[1]{\setcounter{section}{0}
 \setcounter{equation}{0}
 \renewcommand{\theequation}{{#1}.\arabic{equation}}
 \renewcommand{\thesection}{}}

\newcommand{\rf}[1]{(\ref{#1})}
\newcommand{\beqa}[1]{ \hspace{0in}\\        $\scriptscriptstyle{(#1)}$ \vspace{-18pt}         \begin{eqnarray} \label{#1} }

\newcommand{\eeqa}{\end{eqnarray}}
\newcommand{\beq}[1]{ \hspace{0in}\\        $\scriptscriptstyle{(#1)}$ \vspace{-10pt}         \begin{equation} \label{#1} }
\renewcommand{\beq}[1]{\begin{equation} \label{#1} } \renewcommand{\beqa}[1]{\begin{eqnarray} \label{#1} }
\newcommand{\eeq}{\end{equation}}
\newcommand{\ba}[1]{\begin{array}{#1}}
\newcommand{\ea}{\end{array}}
\newcommand{\bosy}[1] { \mbox{\boldmath ${#1}$}}
\newcommand{\eps}{\epsilon}%

\def\singlespacing{\baselineskip=13pt} 
\def\doublespacing{\baselineskip=18pt} 
\begin{document}
\doublespacing
\begin{center}

{\huge Dynamics of thermoelastic thin plates: \\ A comparison of four theories }

\bigskip

Andrew N. Norris

\medskip
Department of Mechanical \& Aerospace
Engineering, \\Rutgers University, 
98 Brett Road, Piscataway, NJ
08854-8058, USA

\end{center}

\begin{center} {\Large Abstract}\end{center}

Four distinct theories describing the flexural motion of thermoelastic thin plates are compared.  The theories are due to Chadwick \cite{Chadwick62b}, Lagnese and Lions \cite{Lagnese88}, Simmonds \cite{Simmonds99} and Norris \cite{Norris05b}.  Chadwick's theory requires a 3D spatial equation for the temperature but is considered the most accurate as the others are derivable from it by different approximations. Attention is given to the damping of flexural waves.  Analytical and quantitative comparisons indicate that the Lagnese and Lions model with a 2D temperature equation captures the essential features of the thermoelastic damping, but contains systematic inaccuracies.  These are attributable to the approximation for the first moment of the temperature used in deriving the Lagnese and Lions equation.  Simmonds' model with an explicit formula for temperature in terms of plate deflection is the simplest of all but is accurate only at low frequency, where the damping is linearly proportional to the frequency.  It is shown that the  Norris model, which is almost as  simple as Simmond's, is as accurate as the more precise but involved theory of Chadwick. 

\medskip \noindent

\vskip 0.5in
\noindent
Short title: Thermoelastic thin plates\\

\medskip\noindent
Contact: \\
norris@rutgers.edu\\ 
ph. 1.732.445.3818\\
fax 1.732.445.3124

\singlespacing

\newpage
\section*{Introduction}
\bigskip
\fontsize{10}{12}\selectfont
\fontsize{12}{14}\selectfont

A recent renewal of  interest in thermoelasticity of vibrating structures is motivated in part by the observation that   thermoelasticity can be the dominant and unavoidable source of dissipation and noise in certain circumstances.  Structures both small and large in size can be effected.    For instance,  carefully controlled experiments have demonstrated that the dominant loss in  silicon micromelectromechanical (MEMS) double paddle oscillators is due to  thermoelasticity  \cite{Photiadis02}.   Conversely, large masses used in devices planned for gravitational wave measurements are  susceptible  to thermoelastic effects, which need to be controlled if the desired sensitivity is to be obtained  \cite{liu00}. The comparative advantages of several related theories for the the dynamics of thermoelastic thin plates are  considered here, with a view towards improved simulations and understanding of thermoelastic damping in devices. 

Although the subject is thin {\it plates}, it should be noted that the picture for thin {\it beams} is well understood. Damping of flexural motion in  beams caused by irreversible thermoelastic coupling is  well described by Zener's \cite{Zener38} theory of 1938.   According to Zener, the  instantaneous or isentropic bending stiffness of the beam relaxes  to the isothermal value  as a result of diffusion of  heat produced by the alternating compression and extension on opposite faces of the thin structure.  Zener correctly surmised that diffusion is predominantly in the through-thickness direction, with little or no lateral (in-plane) heat conduction, although he did not provide a rigorous justification for this assumption.   He also derived the full form of the beam damping in terms of an infinite set of characteristic relaxation times defined by the eigenvalues of the 1-dimensional heat equation across the thickness. Similar and more general results for the beam were obtained by Alblas \cite{Alblas61,Alblas81}, and more recently by Lifshitz and Roukes \cite{Lifshitz00}. 

The situation for thin plates is more complicated by virtue of the two dimensional nature, which introduces the possibility of flexure in two directions.  The goal of researchers has been to derive reduced governing equations which include thermal effects but are similar to the classical Kirchhoff equations.  The first, and to date most successful work in this respect, was Chadwick \cite{Chadwick62b} who provided a more rigorous basis for thermoelastic equations of thin beams  based on the underlying 3-dimensional theory.  The plate equations of  Chadwick  include  3-dimensional variability in the temperature field.  This level of complexity is however unnecessary, as will be made clear, since the temperature variations in the plane of the plate are insignificant in comparison with the diffusion driven temperature gradient across the thickness. 

A distinct theory for thermoelasticity of thin plates was developed by Lagnese and Lions \cite{Lagnese88}.  The Lagnese and Lions theory differs from Chadwick's in that it retains a separate temperature equation, although this equation does not contain variations in the through thickness.  These are eliminated by taking the first moment of temperature across the plate thickness, so that both the temperature and the deflection equation are two-dimensional in space, and coupled.  These equations bear some resemblance to those of Chadwick, and the connection will be discussed further below.   
The equations of Lagnese and Lions have been widely used as the starting point for studies of thermoelastic plates, e.g.  \cite{Lagnese89,Lagnese90,Avalos98,Avalos00,Avalos00b,Avalos04,Rivera97,Rivera98,Rivera98b,Rivera01,Eller00,Eller01,Lasiecka01,Isakov01,Hansen97,Nandakumaran04,Andrews04,Dalsen03,Lebeau98,Giorgi01,Enomoto02,Leiva03,Liu95,Liu97,Kim92,Shibata94,ldt96,Benabdallah98,Benabdallah02,Amendola03,Henry93,Liu97b}.  Many of these works are not concerned with the mechanical applications {\it per se} but are  focused on mathematical control issues that arise from the unique form of the coupled system of partial differential  equations.  Some,  e.g. \cite{Giorgi00}, have derived similar sets of equations  based on the methods used by Lagnese and Lions \cite{Lagnese88}.   

A third theory is due to Simmonds \cite{Simmonds99}, who derived a single  equation governing the plate deflection.  His equation contains a term that acts like a viscous damper but encapsulates the thermoelasticity explicitly.  Simmonds proposed his equation as a simpler alternative to the  coupled equations of Lagnese and Lions, although he did not compare his model with Chadwick's.   

The fourth thin plate theory considered  is the most recent,  due to Norris \cite{Norris05b}.  It is also a single equation for the plate deflection, slightly more complicated than Simmonds', as we will see.  The Norris equation is based on recent  work by  Norris and Photiadis \cite{Norris04} who considered thermoelastic damping in arbitrary structures.  They derived Zener-like results based on the asymptotically small thermal coupling.   They also applied their general theory  to time harmonic oscillation of thin plates and derived a modified plate equation without temperature appearing explicitly;  instead it enters via a frequency dependent viscoelastic term.  

The purpose of this paper is to explore the similarities and differences between these models for thermoelastic flexural motion in thin plates.   No such  study has appeared to date, and the present comparison  is offered in the hope that it provides clarification and ideas to all interested in modeling the coupled thermoelasticity of plates and similar structures.  The emphasis here is on the mechanical accuracy of the various equations, and in order to compare them we consider the most important dynamic feature introduced by the coupling to the temperature field: damping of otherwise nondissipative flexural waves.   The paper begins with an overview of the equations common to all four models in Section \ref{sec1}.  Subsequently, in Section \ref{sec2} the models of Chadwick, Lagnese and Lions, Simmonds and Norris are introduced in a coherent framework, with some new results for the latter three. Section \ref{sec3} contains a parallel analysis of the dispersion relation for quasi-flexural waves according each model, and a detailed discussion and comparison of the damping predicted by each is given in  Section \ref{sec4}.

\section{Common equations} \label{sec1}
The system of equations derived by Chadwick \cite{Chadwick62b} include the other three models as different  approximation.  In order to see this we first describe what they all have in common.  We begin  with the elastic equilibrium equations with no applied forces  
\begin{subequations}
\beqa{e1a}
\sigma_{33,z} + \sigma_{3\beta , \beta} &=& \rho u_{3,tt},
\\
\sigma_{\beta \gamma,\gamma} + \sigma_{3\beta , z} &=& \rho u_{\beta ,tt}, \quad \beta=1,\, 2. 
\label{e1b}
\eeqa
\end{subequations}
The displacement is ${\bf u}(x,y,z,t) = (u_1,\, u_2, \, u_3)$, $\bosy{\sigma}$ is the symmetric stress tensor, $\rho $ is mass density per unit volume, and the repeated Greek suffices indicates summation over $1$ and $2$.  The  plate of thickness $h$ occupies $-\infty < x,\, y < \infty ,\, -h/2 \le z \le h/2$, and is traction free on the surfaces:
$\sigma_{13} = \sigma_{23} = \sigma_{33} = 0$ on $z = \pm \frac{h}{2}$.

Integrating eq. \rf{e1a} with respect to $z$, using the fact that the  shear stresses are constant (zero)  on the surfaces, and that the normal stresses vanish,  implies 
\beq{e5}
 - \int\limits_{-h/2}^{h/2} \, dz \,z \sigma_{3\beta , z \beta}  = \rho h w_{tt} . 
\eeq
where $w$ is the averaged through-thickness displacement
\beq{e4}
 w(x,y,t) = \frac{1}{h} \int\limits_{-h/2}^{h/2} \, dz \, u_3 (x,y,z,t)\, . 
\eeq
The in-plane equilibrium equations \rf{e1b} then imply 
\beq{e6}
M_{\beta \gamma ,\beta \gamma} 
  = \rho h w_{tt} +  \rho \int\limits_{-h/2}^{h/2} \, dz \,z u_{\beta ,  \beta tt} ,  
\eeq
where the first moment of the stresses, or more simply, the moments, are 
\beq{ap5}
M_{\beta \gamma} (x,y,t)= \int\limits_{-h/2}^{h/2} dz  \, z \sigma_{\beta \gamma}, \quad \beta,\,  \gamma =1,2. 
\eeq

We now assume the kinematic {\it ansatz} normally associated with the name  Kirchhoff,  
\beq{e8}
u_\beta  = -z \, w_{, \beta}, \quad \beta =1,2, 
\eeq
so that \rf{e6} becomes
\beq{e9}
M_{\beta \gamma ,\beta \gamma} 
  = \rho h w_{tt}  - \rho I \Delta  w_{tt},  
\eeq
where $I=h^3/12$, and 
$\Delta$ is the in-plane Laplacian, 
$\Delta = \frac{\partial}{\partial x^2} + \frac{\partial}{\partial y^2}$. 
The development so far is independent of the thermoelastic properties of the plate. In fact, equation \rf{e9} is still exact apart from the simplification of the the final term - the rotatory inertia - which is often ignored in a first approximation.  We will do this later, but for the moment retain this inertial term, associated with Rayleigh \cite{Timoshenkotps}. 

The thermal properties enter through the 
 equations for the in-plane stresses in an isotropic thermoelastic plate, which follow   
from eq. (6.5) of \cite{Lagnese88} or elsewhere, 
\beqa{app4}
\sigma_{11} &=& \frac{E}{1-\nu^2} \big( \varepsilon_{11} + \nu\varepsilon_{22}\big) - \frac{\alpha E}{1-\nu}\, \theta \, , \nonumber \\
\sigma_{22} &=& \frac{E}{1-\nu^2} \big( \nu \varepsilon_{11} + \varepsilon_{22}\big) - \frac{\alpha E}{1-\nu}\, \theta\, ,\\
\sigma_{12}&=& \frac{E}{1+\nu}  \varepsilon_{12}\, .   \nonumber 
\eeqa
Here $\varepsilon_{\beta\gamma} = ( u_{\beta ,\gamma} + u_{\gamma ,\beta})/2$ are the in-plane strains,  and $\theta (x,y,z,t)$ is the temperature deviation  from its ambient constant value $T$. Also, $E$ is  Young's modulus, $\nu $  Poisson's ratio,  and    $\alpha$  is the coefficient of thermal expansion.
Combined with eqs. \rf{ap5} through \rf{e9}, these imply the first of our governing equations: 
\beq{e10}
D\Delta^2 w + D\alpha (1+\nu) \Delta  \Theta + \rho h w_{tt}  - \rho I \Delta  w_{tt} = 0\, , 
\eeq
where $D = EI/(1-\nu^2)$ is the bending stiffness, and $\Theta (x,y,t)$ is the (normalized) first moment of temperature: 
\beq{ch2} 
\Theta (x,y,t) = \frac{1}{I}\, 
\int\limits_{-h/2}^{h/2} dz  \, z \, \theta (x,y,z,t)\, . 
\eeq

Equation \rf{e10} is the standard plate equation with the added term $\Theta$ which couples the plate deflection to the thermal effects.  The four theories all use  equation \rf{e10} in one form or another, although an additional  equation is required to close the system.   The theories considered differ  in how the temperature moment $\Theta$ is calculated in each.  Thus, Simmonds \cite{Simmonds99} and Norris \cite{Norris05b} give explicit though different forms for $\Theta$.   In the theory of Lagnese and Lions \cite{Lagnese88} the function $\Theta (x,y,t)$ satisfies a 2+1 dimensional partial differential equation. 
Chadwick's model  \cite{Chadwick62b} requires solving a 3+1 dimensional equation for $\theta (x,y,z,t)$, and as such, is the most general of the theories.  We begin with this. 

\section{The four theories} \label{sec2}

\subsection{Chadwick's thermoelastic flexural plate equations}
Chadwick \cite{Chadwick62b} derived a coupled system of equations for $w(x,y,t)$ and $\theta (x,y,z,t)$.  We begin with  the relation for entropy $s$ in terms of strain $\bosy{\varepsilon}$ and temperature deviation from  ambient,
\beq{app1}
s = \frac{E\alpha}{1-2\nu} \varepsilon_{kk} + \frac{C_{}}{T}\theta,
\eeq
and the heat flux condition
\beq{app2}
Ts_t = K\big( \Delta \theta + \theta_{zz}\big) . 
\eeq
Here,  $C$ and $K$ are the heat capacity and thermal conductivity, respectively. 
Equations \rf{app1} and \rf{app2}  are well known, e.g. \cite{Nowacki}, and are eqs. (6.16) and (6.19) of \cite{Lagnese88}. 
Under the thin plate assumption, the stress $\sigma_{33}$ vanishes and therefore 
\beq{app0}
s = \frac{E\alpha}{1-\nu} (\varepsilon_{11}+\varepsilon_{22}) + \frac{C_{}}{T}\theta \, . 
\eeq
 Substituting from this into \rf{app1} and \rf{app2},  combined with the Kirchhoff assumption \rf{e8} yields 
\beq{app3}
C_{}\theta_t- K( \Delta \theta + \theta_{zz}\big)  - \frac{z\alpha TE}{ 1-\nu}\Delta w_t = 0.   
\eeq
This is the second equation of Chadwick, which, together with  the deflection equation \rf{e10}, forms a closed pair of partial differential equations.  The  remaining ingredients are boundary conditions; however, since we consider the plate of infinite lateral extent, no edge conditions are necessary.  The PDEs \rf{e10} and \rf{app3} need to be supplemented by conditions for  the temperature at the surfaces. 

In Chadwick's analysis, and in the later comparisons,  the rotatory inertia in \rf{e10} is ignored. In addition we consider the simple no-flux conditions for temperature on the plate surfaces.  With these clarifications,  the Chadwick model may be summarised as 
\begin{subequations}\label{ch}
\beqa{chs}
D\Delta^2 w + D\alpha (1+\nu) \Delta  \Theta + \rho h w_{tt} &=&0, 
\\
C_{}\theta_t - K (\Delta \theta + \theta_{zz}) - \frac{z\alpha TE}{1-\nu} \, \Delta  w_t &=&0, \label{chsb}
\\
\theta_z(x,y,\pm \frac{h}{2},t)&=& 0. \label{chsc}
\eeqa
\end{subequations}

Chadwick's  derivation included additional details which gave careful account of whether  quantities are isotropic or isentropic\footnote{Chadwick employed a different definition of the thermal expansion coefficient $\alpha$, i.e. $1/3$ of the quantity  used here.}.   We ignore this distinction here but refer readers to the original paper for a proper description of the governing equations.

\subsection{The Lagnese and Lions model}\label{subsec1}

The Lagnese and Lions \cite{Lagnese88}  equations for  flexural-thermal dynamics of a thin plate are 
\begin{subequations}
\beqa{ch4}
D\Delta^2 w + \frac{D}{2}\alpha (1+\nu) \Delta  \Theta + \rho h w_{tt}  - \rho I \Delta  w_{tt}  &=&0, \\ 
C_{}\Theta_t - K \Delta  \Theta + \frac{h}{I}K \Theta - \frac{\alpha TE}{1-2\nu} \, \Delta  w_t &=&0.  \label{ch5}
\eeqa
\end{subequations} 
These equations  follow from, for instance,  eq. (6.32) of \cite{Lagnese88} with zero forcing ($p=f_3=0$ in \cite{Lagnese88}) and with zero-flux boundary conditions on the top and bottom faces of the plate $(\lambda_1 = 0$ in \cite{Lagnese88}).  The first equation \rf{ch4} is a modification of the standard equation of flexural motion for a thin plate , while eq. \rf{ch5}  for the temperature moment, $\Theta (x,y,t)$, is specialized to  
the case of zero thermal flux on the faces $z=\pm h/2$.  Lagnese and Lions \cite{Lagnese88} included the possibility of more general flux conditions. The final term in \rf{ch4} represents rotatory inertia, analogous to Rayleigh's theory, and is a common addition to the classical theory \cite{Timoshenkotps}.  

The equations as derived \cite{Lagnese88} contain two   errors which need to be addressed before the model can be compared with others.  First, the factor of $1/2$ in the second term of \rf{ch4} is incorrect, and should be unity.  The error is a result of   improper application of the variational derivation of the equation.  The error is evident by comparison of \rf{ch4} with \rf{e10}.  Appendix A provides  details on the necessary correction to the variational formulation. 
The second error is the  factor $(1-2\nu)^{-1}$ in the final term of \rf{ch5}, which   should be $(1-\nu)^{-1}$. The error arises from  failure to properly apply the entropy constitutive relation to the thin plate deformation as done in eqs. \rf{app1} and \rf{app0} (see eq. 6.19 of \cite{Lagnese88}).  

Equation \rf{ch5} may be obtained as the first moment of eq. 
\rf{app3} divided by $I$. The  term $(h/I)K \Theta $ in eq. \rf{ch5} 
comes from the   evaluation of $(K/I)\mu $ where $\mu (x,y,t)$ is the first moment of $ - \theta_{zz}$ \cite{Lagnese88}.  Applying the assumed zero flux conditions on the plate faces \rf{chsc}, the moment reduces to, with no approximation,
\beqa{th2}
\mu &\equiv & - \int\limits_{-h/2}^{h/2} dz  \, z \theta_{zz}
\nonumber \\
&=& \theta(x,y,\frac{h}{2},t)- \theta(x,y,-\frac{h}{2},t)\, .
\eeqa
In order to relate this to $\Theta$,   Lagnese and Lions assume  that $\theta (x,y,z,t)$ is linear in $z$ (even though this implies that $\theta_{zz} \equiv 0$), i.e. 
\beq{lin}
\theta (x,y,z,t) \approx z\Theta (x,y,t),\quad \Rightarrow \quad  \mu = h\Theta , 
\eeq
and hence the term  $(h/I)K \Theta $ in eq. \rf{ch5}.

For the sake of simplicity we choose to ignore the rotatory inertia term $\rho I \Delta  w_{tt}$ in order to compare the 
Lagnese and Lions equations with the other thermoelastic thin plate theories. 
which is not central to the  Lagnese and Lions model, and is normally ignored in other first order thin plate models.  This term could  be included but presents algebraic complications that make comparisons between models unnecessarily clumsy.  With this said, and 
  with the  corrections noted above,  the following system  represents the Lagnese and Lions model for further discussion:
\begin{subequations}\label{ll}
\beqa{llz}
D\Delta^2 w + D\alpha (1+\nu) \Delta  \Theta + \rho h w_{tt}    &=&0, \\ 
C_{}\Theta_t - K \Delta  \Theta + \frac{h}{I}K \Theta - \frac{\alpha TE}{1-\nu} \, \Delta  w_t &=&0.   \label{llzb}
\eeqa
\end{subequations}

\subsection{Simmonds' model}

Simmonds proposed his model as a simpler alternative to the Lagnese and Lions equations.  The final result is a single governing equation, which we rederive here using slightly different methods. Simmonds' analysis is premised on certain assumed scalings, which for our purposes are unnecessary as we will retain the leading order terms by physical argument. 
In the process of his derivation, Simmonds  calculates the shear stresses in the plate. Since this is in itself a useful result, we begin there.    Thus, \rf{e1b} and the in-plane stresses of \rf{app4},  imply 
\beq{sim1}
\sigma_{3\beta , z} = - \frac{E}{1-\nu^2} \big[ \frac{1}{2} (1+\nu)u_{\gamma , \gamma \beta  } 
+ \frac{1}{2} (1-\nu) u_{\beta  , \gamma  \gamma  } -  (1+\nu) \alpha \theta_{,\beta}\big] 
+ \rho u_{\beta ,tt}\, . 
\eeq
This is comparable to eq. (3.5) of \cite{Simmonds99}, although the latter does not include the final term in \rf{sim1}, which we retain in order to be consistent with the general derivation of \rf{e10}. 
This term is absent in Simmonds' analysis by virtue of the scalings chosen.  Using the Kirchhoff assumption of \rf{e8}, 
\beq{s2}
\sigma_{3\beta , z} =  \frac{E}{1-\nu^2} \big[ z\Delta w_{,\beta}+  (1+\nu) \alpha \theta_{,\beta}\big] 
- \rho z w_{,\beta tt}\, ,  
\eeq
which can be integrated,  to yield an explicit equation for the shear stresses in terms of $w$ and $\theta$,
\beq{s3}
\sigma_{3\beta } =  \frac12 (z^2 - \frac{h^2}{4}) \big( \frac{E}{1-\nu^2}  \Delta w_{,\beta} - \rho  w_{,\beta tt} \big)
+  \frac{E \alpha}{1-\nu}  
\int\limits_{-h/2}^{z} dz'  \, 
\theta_{,\beta} (x,y,z',t).   
\eeq
This clearly satisfies the conditions that the shear stresses vanish on the surfaces as long as $\theta (x,y,z,t)$ is an odd function of $z$, which is assumed.  Equation \rf{s3} is analogous to (4.11) of \cite{Simmonds99} although  the latter contains an explicit form for the integral of $\theta$.  We will return to this point in detail later, and leave \rf{s3} in its general form at this stage.  Substituting from \rf{s3}   gives
\beq{s4}
D\Delta^2 w - \rho I \Delta  w_{tt} - \frac{E \alpha}{1-\nu}  \int\limits_{-h/2}^{h/2} dz  \,
\int\limits_{-h/2}^{z} dz'  \, 
\Delta \theta  (x,y,z',t)
+ \rho h w_{tt}   = 0\, ,   
\eeq
and integrating by parts reproduces eq. \rf{e10} exactly.  The Simmonds model therefore includes the basic equation for the plate deflection.  It remains to discuss his equation for the temperature. 

Without going into the details of the  scaling employed, it suffices to note that Simmonds solves eqs. \rf{chsb} and \rf{chsc} with $\Delta \theta \rightarrow 0$, i.e. 
\begin{subequations}
\beqa{s5}
   \theta_{zz} - \frac{C}{K}\theta_t  +\frac{z\alpha TE}{K(1-\nu  )} \, \Delta  w_t &=&0 ,  
\\
\theta_z(x,y,\pm \frac{h}{2},t)&=& 0.  \label{s5b}
\eeqa
\end{subequations}    
These are equations (4.3) and (4.4) of \cite{Simmonds99}. 
Simmonds' solution may be expressed in a slightly different form than the original (eq. (4.9) of \cite{Simmonds99}).
Consider the {\it ansatz} 
\beq{x1}
\theta (x,y,z,t) = \sum\limits_{n=1}^\infty u^{(n)}(\frac{2z}{h})\, v^{(n)}(x,y,t) + \theta_H,
\eeq
where the series represents the particular solution, and $\theta_H$ is a solution of the homogeneous equations which is required in order to give the correct initial condition for the temperature. The latter may be expressed in terms of eigenfunctions, as in \cite{Simmonds99}. 
The functions $u^{(n)}(z)$ are defined  such that:
\beq{x2}
{u^{(n)}} ''(s) = u^{(n-1)}(s), \quad  {u^{(n)}} '(\pm 1) = 0, \quad n\ge 1 ;\quad u^{(0)}(s) \equiv s.
\eeq
These may be obtained recursively as described in Appendix B.  Note that $\theta$ of \rf{x1} automatically satisfies the zero flux conditions on the surfaces.

Substituting from \rf{x1} into \rf{s5} and equating the coefficients of $u^{(n)}$ to zero implies 
recursion relations for $v^{(n)}$,
\beq{x3}
v^{(n+1)}(x,y,t)   = \frac{h^2 C}{4K}\, v^{(n)}_t (x,y,t), \, \, n\ge 1,\qquad
v^{(1)}(x,y,t) =  - \frac{h^3 \alpha TE}{8K(1-\nu  )} \, \Delta  w_t  \, . 
\eeq
Thus, formally at least, 
\beq{x12}
\theta (x,y,z,t) =  -\frac{h\alpha TE}{2C(1-\nu  )} \,  \sum\limits_{n=1}^\infty u^{(n)}(\frac{2z}{h})\, \big(\tau_S \frac{\partial}{\partial t}\big)^n\, \Delta  w\, +  \theta_H. 
\eeq
where 
\beq{x31}
\tau_S  = \frac{h^2 C}{4K} , 
\eeq
is a characteristic time for thermal diffusion across the plate thickness\footnote{Note that $\tau_S = \frac{\pi^2 }{4 }\, \tau_0$, where $\tau_0$,  defined in eq. \rf{tau}, is the fundamental time scale for all the models.}.  The first moment of \rf{x12} can be evaluated using eq. \rf{x2}, with the result
\beqa{s61}
  \Theta (x,y,t)  &=& 
  \frac{3\alpha TE}{C(1-\nu  )} \,  
 \sum\limits_{n=1}^\infty u^{(n+1)}(1)\, \big(\tau_S \frac{\partial}{\partial t}\big)^n\, \Delta  w\, +  \Theta_H, 
 \nonumber \\
 &=& \frac{\alpha TE}{C(1-\nu  )} \,\bigg[ 
 \frac25 \tau_S \frac{\partial}{\partial t}  - \frac{17}{105} \big(\tau_S \frac{\partial}{\partial t}\big)^2 + 
 \frac{62}{945} \big(\tau_S \frac{\partial}{\partial t}\big)^2 + \ldots \bigg]\Delta  w\,  + \Theta_H,
\eeqa
with obvious meaning for $\Theta_H$, which we assume is zero for simplicity. The  coefficients $u^{(n+1)}(1)$ follow from the general formula \rf{gf} in Appendix B.   The convergence properties of the series expansion are also discussed in Appendix B. 

Simmonds argues that the times of interest are much longer than $\tau_S$.  Therefore only the first term in the expansion of the particular solution is significant, yielding 
 an explicit expression for the first moment, 
\beq{s8}
  \Theta  = \frac{h^2\alpha TE}{10 K(1-\nu)} \, \Delta  w_t\, ,   \quad \mbox{Simmonds}.  
\eeq
This implies a  single governing equation for the plate deflection (eq. (4.13) of \cite{Simmonds99})
\beq{s11}
D\Delta^2 w + \eps_\nu \frac{2}{5} \tau_S D\Delta^2 w_t  + \rho h w_{tt}   = 0\, ,   
\eeq
where the nondimensional factor $\eps_\nu$ is defined as
\beq{epsnu}
\epsilon_\nu = \frac{ \alpha^2 TE}{C_{}}\big(\frac{1+\nu}{1-\nu}\big)\, . 
\eeq
We will use this quantity repeatedly in the remainder of the paper. 

It is interesting to note that of all the theories Simmonds' does not involve the heat capacity $C$ in the final equation.  This is a consequence of the fact that the approximation \rf{s6}  is equivalent to ignoring the time derivative of $\theta$ in \rf{s5} and solving instead
\beq{s7}
  K  \theta_{zz}  = - \frac{z\alpha TE}{1-\nu} \, \Delta  w_t\, ,    
\eeq
subject to the zero-flux conditions \rf{chsc}.  The solution of this is obviously independent of the heat capacity, and is  
\beqa{s6}
  \theta (x,y,z,t)  = \frac{z}{2}\big( \frac{h^2}{4}- \frac{z^2}{3}\big)\,  \frac{\alpha TE}{K(1-\nu)} \, \Delta  w_t\,  .
\eeqa
Substituting this approximation for the temperature distribution into \rf{ch2} yields \rf{s8} directly. 

\subsection{Norris' thermoelastic flexural plate equation}

Norris' model is a single viscoelastic-type equation for plate deflection $w(x,y,t)$, 
\beq{n1}
(1+\eps_\nu )\, D\Delta^2 w - \eps_\nu\, D\, g*\Delta^2 w
 + \rho h w_{tt} = 0 ,
\eeq
where $*$ denotes convolution in time and $\eps_\nu$ is defined in \rf{epsnu}. The relaxation function $g(t)$ is 
\beq{f0}
g(t) = \sum\limits_{n=0}^\infty  \,\frac{d_n}{\tau_n}  \, e^{-t/\tau_n},  
\eeq
with 
\beq{mod1}
d_n = \frac{96}{\pi^4 (2n+1)^4} ,\quad 
\tau_n = \frac{\tau_0}{(2n+1)^2} ,\quad n=0,1,2,\ldots ,  
\eeq
and 
\beq{tau}
\tau_0 = \frac{h^2 C_{}}{\pi^2 K}. 
\eeq

The single equation \rf{n1} may be obtained in a variety of ways.  For instance, it is clear from Appendix B that it is equivalent to retaining the complete series in eq. \rf{s61} rather than just the first term as Simmonds does.  A more direct derivation is outlined in Norris and Photiadis \cite{Norris04} in which the time-harmonic version of \rf{n1} is derived.  Briefly, this follows by solving \rf{s5}  in the frequency domain and converting back to the time domain (see  eqs. \rf{ap1} and \rf{bb2}) yielding
\beq{ch3} 
\Theta  =  \frac{\alpha TE}{C_{}(1-\nu)} \, \big( \Delta  w - g* \Delta  w \big)\, . 
\eeq
When combined with \rf{chs}, this reduces the plate model to eq. \rf{n1} alone.  The justification for ignoring the in-plane spatial derivatives in \rf{chsb} is motivated by clear physical arguments in  \cite{Norris04} and is also evident from the results below where we will see that predictions of the Chadwick and Norris models are essentially identical.  See \cite{Norris05b} for a more thorough and complete derivation of the simplified thermoelastic plate theory, along with appropriate plate boundary conditions and example applications. 

\subsection{Summary of the four models}

All models share eq. \rf{e10}, which we simplify for present purposes by ignoring the rotatory inertia, so that 
\beq{4m1}
D\Delta^2 w + D\alpha (1+\nu) \Delta  \Theta + \rho h w_{tt}  = 0\, .  
\eeq
The theories differ in how they estimate $\Theta$ of \rf{ch2}.  Thus, 
\begin{subequations}
\beqa{4m2}
&&C_{}\theta_t - K (\Delta \theta + \theta_{zz}) - \frac{z\alpha TE}{1-\nu} \, \Delta  w_t =0,\quad \mbox{with  }
\theta_z(x,y,\pm \frac{h}{2},t)= 0, \quad \mbox{Chadwick},\, \, 
\\
&&C_{}\Theta_t - K \Delta  \Theta + \frac{h}{I}K \Theta - \frac{\alpha TE}{1-\nu} \, \Delta  w_t =0, \qquad \mbox{Lagnese \& Lions},
\\
&& \Theta =  \frac{\eps_\nu}{\alpha(1+\nu) } \, \times \,  \left\{
\ba {lc}
\frac{\pi^2}{10} \tau_0  \Delta w_t ,& \qquad \quad \mbox{Simmonds}, \\
 & \\
 (  \Delta w - g* \Delta w) ,& \quad \mbox{Norris}. 
 \ea \right.
\eeqa
\end{subequations}
In Chadwick's model $\Theta$ is determined indirectly by first solving the equation for $\theta (x,y,z,t)$, while $\Theta (x,y,t)$ is found directly as the solution of a PDE in Lagnese and Lions' model.  The Simmonds and Norris models provide explicit expressions for  $\Theta$, although in the latter case a convolution is required. We now look in detail at solutions to these equations. 

\section{Flexural waves and attenuation} \label{sec3}

\subsection{Traveling wave solutions}

In order to compare the four models we disregard any boundary effects and consider the infinite plate.  The appropriate fundamental solutions are traveling time harmonic straight crested waves.  We examine the form of these waves for the different theories with a view towards distinguishing the models and determining a best approximation to the underlying physics. 
The assumed form of the solution  is 
\begin{subequations}
\beqa{twe}
\big\{ w(x,y,t),\, \theta (x,y,z,t) \big\} &=& \big\{w_0,\, \theta_0 u(z) \big\}  \, e^{i(k x - \omega t)},  \qquad \mbox{Chadwick}, 
\\
\big\{ w(x,y,t),\, \Theta (x,y,t) \big\} &=& \big\{w_0,\, \Theta_0  \big\}  \, e^{i(k x - \omega t)},  \qquad \mbox{Lagnese \& Lions},
\\
w(x,y,t) &=& w_0 \, e^{i(k x - \omega t)} ,  \qquad \qquad \mbox{Simmonds, Norris}. 
\eeqa
\end{subequations}
The frequency $\omega$ is assumed to be real and positive. 
Define the real-valued flexural wavenumber $\kappa >0$ and the complex-valued thermal wavenumber $\gamma$ by 
\beq{defk}
\kappa^4  = \omega^2 \frac{\rho h }{D},\qquad 
\gamma^2  = i\omega \frac{C_{}}{K}. 
\eeq
The dispersion relation between $k$ and $ \omega$    has a common form for in all four  plate theories
\beq{disp}
k^4  \bigg[
1 +  \eps_\nu   (\gamma h)^2\,  F
\bigg] - \kappa^4 = 0 ,  
\eeq
where  $F$ is a different quantity  for each,
\beqa{funf}
F = \left\{
\ba{ll}
F_C\big( (\gamma^2 - k^2)^{1/2}  h\big) , &  \mbox{Chadwick}, \\
 & \\
 F_L\big( (\gamma^2 - k^2)^{1/2} h\big), &  \mbox{Lagnese and Lions},\\
& \\
-\frac{1}{10}  , &  \mbox{Simmonds},  \\
& \\
F_C\big( \gamma h\big) , &  \mbox{Norris},
 \ea
 \right.
 \eeqa
and the functions $F_C$ and $F_L$ are 
\beq{fl}
F_C(x) =
\frac{1}{x^2} + \frac{24}{x^5}\big( \frac{x}{2} - \tan \frac{x}{2} \big), \qquad  
F_L(x) = \frac{1}{x^2 -12} .  
 \eeq
These results follow from the identity  \cite{Norris04} 
\beq{ap1}
f(\gamma h) = 1- \widetilde{g}(\omega) , 
\quad \mbox{where}\quad
\widetilde{g}(\omega)=   \sum\limits_{n=0}^\infty  \,\frac{d_n}{1 - i\omega \tau_n } , 
\eeq
and the  $\widetilde{g}(\omega) $ is the Fourier transform of $g(t)$.

All of the  models considered are deviations from the non-dissipative elastic Kirchhoff model of thin plate dynamics. Any changes in wave speed, energy propagation speed, real wave number, modal frequency, or a quantity associated  with  energy preserving systems, is expected to be a perturbation of the value without thermal effects.   It makes sense to consider and compare the effects of the different models on the attenuation as this is an order unity effect. 
While there are several measures of dissipation, we focus here on the   quality factor, $Q$,  a non-dimensional parameter, defined in the present context by the imaginary part of the wavenumber as
\beq{q}
Q^{-1} = 4\,\mbox{Im} \, \big(\frac{k}{\kappa} \big).  
\eeq
The  quality factor can be related to, for instance the attenuation per unit length of a  
 propagating wave as $\omega /(2 v Q) $, where $v$ is 
the  speed of energy propagation.

\subsection{Perturbation expansion}

The dispersion relation can be expressed as a cubic in $k^2$ for the Lagnese and Lions model.  However, rather than solve this we take advantage of the fact that for materials of interest $\eps_\nu \ll 1$, see Table I.   This permits solution of \rf{disp} by regular perturbation.  The leading order roots correspond to the classical theory, i.e. $k^4 - \kappa^4=0$ and we consider the propagating root with unperturbed solution $k=\kappa$.   The leading order solution is therefore the  non-dissipative flexural wave with real phase speed $\omega/\kappa$ and real energy propagation speed  $2\omega/\kappa$.     The first order term in the perturbed solution is given by  
\beq{disp1}
\frac{k}{\kappa}  = 
1 -  \frac{\eps_\nu}{4}  (\gamma h)^2\,  F_0
 + \mbox{O}(\eps_\nu^2) ,  
\eeq
where  $F_0$ for the  different models is  
\beqa{funf1}
F_0 = \left\{
\ba{ll}
F_C\big( (\gamma^2 - \kappa^2)^{1/2}  h\big) , &  \mbox{Chadwick}, \\
 & \\
F_L\big( (\gamma^2 - \kappa^2)^{1/2} h\big), &  \mbox{Lagnese and Lions},\\
& \\
-\frac{1}{10}  , &  \mbox{Simmonds} ,\\ 
& \\
F_C\big( \gamma h\big) , &  \mbox{Norris}.
 \ea
 \right.
 \eeqa

\bigskip
\begin{center}

{\small
\begin{tabular}{|l|c|l|} \hline  \hline
 & & \\ 
  Material   &     $\epsilon_\nu \, \times\, 10^3$    & \, \, $ l$   \\  
            &          &  (nm)
 \\ \hline 
 & &\\
Aluminum 	& 8.38 & 32.3 \\
Beryllium & 5.91 & 11.2	\\
Copper 		& 5.71 & 91.8\\
Diamond 	& 0.13 & 58.1\\
GaAs 			& 0.96 & 230 \\
GaPh 			& 0.69 & 38.1 \\
Germanium & 1.08 & 25.7 \\
 Gold 		& 0.42 & 143\\
 Lead 		& 19.11 & 54.1\\
Nickel 		& 5.15 & 13.6\\
Platinum 	& 2.92 & 24.9\\
Silicon 	& 0.40 & 28.1\\
SiC 			& 0.87 & 5.81\\
Silver 		& 7.50 & 177 \\
 & &\\
\hline
\end{tabular}
}

\medskip 
 Table I. The thermoelastic coupling parameter  of  eq. \rf{epsnu}   and the phonon mean free path $ l$ of eq. $\rf{hs}_1$, both at at 300 K.  
\end{center}

The replacement  $k\rightarrow \kappa$ in \rf{funf1} is important as it leads to further significant 
simplification, as we see next. Specifically, eq.  \rf{disp1} may be replaced by 
\beq{dp}
\frac{k}{\kappa}  = 
1 -  \frac{\eps_\nu}{4}  \,  f ( \gamma h)
 + \mbox{O}(\eps_\nu^2,\, \eps_\nu \delta) ,  
\eeq
where $\delta \ll 1$ is defined below, and the function   $f$ is 
\beqa{f1}
f(x) = \left\{ \ba{lc}
1 + \frac{24}{x^3}\big( \frac{x}{2} - \tan \frac{x}{2} \big), & \quad \mbox{Chadwick, Norris},  
\\ & \\
\frac{x^2}{x^2 -12}, &  \quad \mbox{Lagnese and Lions}  
\\ & \\ 
-\frac{x^2}{10}, &  \quad \mbox{Simmonds}.
\ea \right. 
 \eeqa
First note that 
\beq{ss}
\frac{\kappa^2}{\gamma^2} =  -ia \delta , \qquad \mbox{where}\quad \delta \equiv \frac{l}{h}, 
\eeq
and  the length $l$ and  nondimensional parameter $a$  are defined as 
\beq{hs} 
l  = \frac{3K }{\bar c \, C_{} } \ , \qquad
a = \frac{2}{\sqrt{27}}\big[\sqrt{2(1-\nu)}  + \frac{(1-\nu)}{\sqrt{1-2\nu}}] \, . 
\eeq 
Here,  $\bar c$ is the averaged elastic wave speed, $\bar{c} = (c_L+2c_T)/3$, where 
$c_T^2 = E/[2(1+\nu)\rho]$, $c_L^2 = E(1-\nu)/[(1+\nu)(1-2\nu) \rho]$.   The length $l$ is defined in this way so that it  is identifiable as a thermal phonon mean free path, since  $\bar c$ is a typical  phonon  speed \cite{Kittel}.  The parameter $a$ 
is of order unity.   Values of $l$ for a variety of materials are given in Table 1.  The phonon mean free path  must be far less than a typical structural length scale if 
 thermoelasticity  theory is to remain valid, and in particular we must have $l \ll h$.  Consequently, 
 \beq{smd}
 \delta \ll 1, 
 \eeq
  and  
\beq{sss}
(\gamma^2 - \kappa^2)^{1/2} h =   \gamma h \, \big[ 1 + \mbox{O}(\delta )\big]\ . 
\eeq
This provides the justification for the simplification \rf{dp} and \rf{f1}.  Alternative arguments, not based on the phonon mean free path, but which give the same result are given in Appendix C. 

\subsection{The attenuation }

The leading order approximation to the damping of the thermoelastic plate wave is, based  upon \rf{q}, \rf{disp1} and the approximation for $F_0$, 
\beq{disp4}
Q^{-1}  = 
  -  \eps_\nu  \,   \mbox{Im}\, f(\gamma h) . 
\eeq

The quantity $f(\gamma h)$ is simple for the Lagnese \&  Lions model and Simmonds' model, for which \rf{disp4} is exactly
\beqa{disp23}
Q^{-1}  =
  \eps_\nu   \, \times \, \left\{ \ba{lc}
  \frac{\omega \tau_L }{1+ \omega^2 \tau_L^2}, &   \quad \mbox{Lagnese and Lions}, 
  \\ & \\
  \omega \tau_0 \, \frac{\pi^2 }{10} , & 
  \quad \mbox{Simmonds}, 
  \ea\right.
\eeqa
where
\beq{tau1}
\tau_L  =  \frac{\pi^2 }{12}\, \tau_0 . 
\eeq

Turning to the damping as predicted by Chadwick's model, which is the same as for Norris' by eq. \rf{f1}, eqs. \rf{disp4} and \rf{ap1} imply 
\beq{dispc}
Q^{-1}  = 
  \eps_\nu   \sum\limits_{n=0}^\infty  \,   \frac{d_n \omega \tau_n }{1+ \omega^2 \tau_n^2}, 
  \quad \mbox{Chadwick, Norris}. 
\eeq

\section{Comparing the four models}  \label{sec4}

\subsection{Comparison of attenuation}

 In order to compare the formulae for the damping, define
\beq{s1}
S = \sum\limits_{n=0}^\infty  \,   \frac{d_n \omega \tau_n }{1+ \omega^2 \tau_n^2}, 
\qquad
S_L = \frac{ \omega \tau_L }{1+ \omega^2 \tau_L^2}\, ,
\qquad
S_S =  \omega \tau_0 \, \frac{\pi^2 }{10} , 
\qquad
S_0 =  \frac{ \omega \tau_0 }{1+ \omega^2 \tau_0^2}\, . 
\eeq
Thus, $Q^{-1}$ is equal to $\eps_\nu  S$ for the Chadwick and Norris models, $\eps_\nu  S_L$ for Lagnese \& Lions, and 
$\eps_\nu  S_S$ for Simmonds' model, respectively.  The single term expression $S_0$ will be explained presently. 

Of the four predictions, Simmonds' is the only one which gives damping that grows without bound as $\omega \rightarrow \infty$.  The Simmonds equation is not of the form of a standard relaxation, but more like a pure viscous damping effect.   We note however, that at low frequency, $S = S_S + $O$\big( (\omega\tau_0)^3\big)$, which follows from 
the identity $\sum\limits_{n=0}^\infty d_n \tau_n = \tau_0 \pi^2 /10$.  The  model predicts the right linear dependence as $\omega \tau_0 \rightarrow 0$. In summary, Simmonds's model is correct at low frequency only. 

\begin{figure}[htbp]
				\begin{center}	
		\includegraphics[width=4.0in , height=2.8in 					]{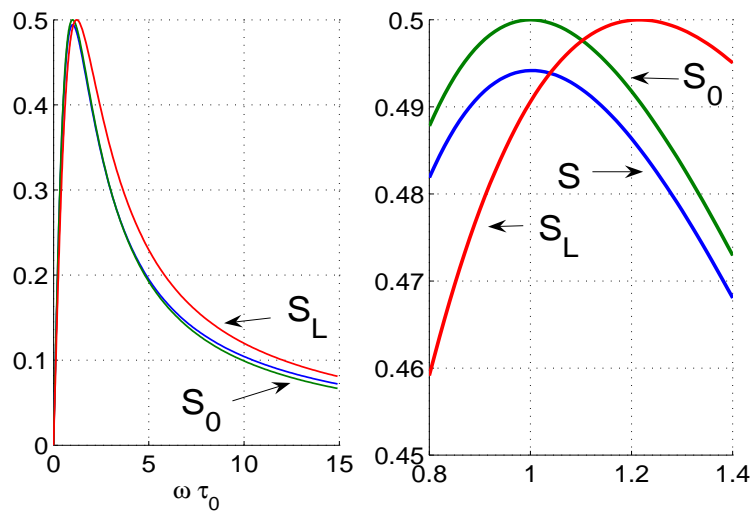} 
		\caption{\small The functions $S$, $S_L$, and $S_0$ of eq. \rf{s1}.	The right hand frame focuses on the peaks near $\omega \tau_0 = 1$.}
		\label{fx1} \end{center}  
	\end{figure}

The three expressions $S$, $S_L$, and $S_0$ are compared in Figure \ref{fx1} as functions of $\omega\tau_0$. 
It is evident from Figure \ref{fx1} that while the Lagnese and Lions function $S_L$ has the same overall behavior as the more precise prediction $S$, the latter is better approximated by $S_0$.  
The maximum 
{\it absolute} error in approximating the sum $S$ by $S_0$ is less than $6\times 10^{-3}$.  By comparison, the absolute error obtained by approximating  $S$ with  $S_L$ is on the order of 10 times larger.  More serious than this, perhaps, is the fact that  the frequency at which the maximum of $S_L$ occurs is noticeably in error,  at $\omega \tau_0 = 12/\pi^2 = 1.216$, see Fig. \ref{fx1}. 

It is interesting to note that the maximum of $S$ occurs close to but not exactly at $\omega \tau_0 = 1$. The peak occurs when 
\beq{smax}
\frac{S}{ \omega \tau_0 } = \frac{\pi^2}{20} = 0.4935 .  
\eeq
This must be solved numerically for $\omega\tau_0$, which can be done using with \rf{s1} or the following alternative formula for $S$, which follows from \rf{f1}, and from  \cite{Norris04}, 
\beq{salt}
S= \frac{6}{x^3}\, \bigg[ x - \bigg( \frac{ \sin x + \sinh x}{\cos x + \cosh x}  \bigg) \bigg] 
\, , \qquad x = \frac{\pi}{\sqrt{2}}\, (\omega \tau_0)^{1/2}\, . 
\eeq
We find that the maximum is at $\omega\tau_0 = 1.0014143\ldots$, and the corresponding maximum value is $S_{\rm max}= 0.494178\ldots$.  The approximant $S_0$ has its maximum precisely at $\omega \tau_0 = 1$, and its maximum value $(1/2)$ is slightly in excess of the correct value $S_{\rm max}$ by about 1\%. 

\subsection{Discussion}

We can gain further understanding of  the above results  by returning to the governing plate equations. 
The approximation \rf{sss} may  be interpreted in terms of the in-plane spatial derivatives in both \rf{llzb}  and \rf{chsb}.  Any variations with wavenumber on the order of $\kappa$ or longer can be safely ignored because  the thermal diffusion makes the variation in $z$ dominant.   Therefore, for flexural waves and most disturbances associated with thin plate dynamics, we can ignore the $\Delta \Theta $ in \rf{llzb} and $\Delta \theta $ in \rf{chsb}.  With the latter approximation, eqs. \rf{chsb} and \rf{chsc} can be solved to yield eq. \rf{ch3}, as discussed before.  In this way the Norris model and the Chadwick model give essentially identical predictions. It also indicates that the Chadwick plate model, while correct, is unnecessarily cumbersome: the in-plane spatial derivatives of temperature can be safely  ignored \cite{Norris04}.     
By the same argument, we may  ignore the term $K \Delta  \Theta$ in \rf{llzb}, so that the correspondingly  simplified form of the Lagnese and Lions model is 
\beq{llc}
C_{}\Theta_t  + \frac{12}{h^2}K \Theta = \frac{\alpha TE}{1-\nu } \, \Delta  w_t .   
\eeq
This can be solved as
\beq{ll4} 
\Theta  =  \frac{\alpha TE}{C_{}(1-\nu)} \, \big( \Delta  w - g_L* \Delta  w \big)\, , 
\eeq
where
\beq{GL}
g_L(t) = \frac{1}{\tau_L}\, e^{-t/\tau_L}\, . 
\eeq
Substituting from \rf{ll4} into \rf{llz} it follows that  the Lagnese and Lions thermoelastic plate model reduces to 
\beq{LLN}
(1+\eps_\nu )\, D\Delta^2 w - \eps_\nu\, D\, g_L*\Delta^2 w
 + \rho h w_{tt} = 0 .
\eeq

\begin{figure}[htbp]
				\begin{center}	
			\includegraphics[width=4.0in , height=3.2in 					]{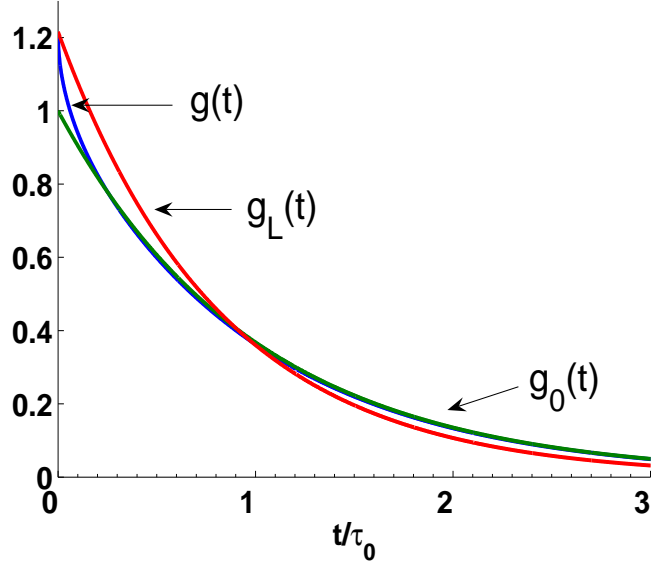} 
		\caption{\small The function $\tau_0 g(t)$ compared with   $\tau_0 g_0(t)$ and $\tau_0 g_L(t)$ .	}
		\label{fx2} \end{center}  
	\end{figure}

Referring to eqs. \rf{f0} and \rf{s1}, it is clear that the one term damping function $S_L$ can be associated with the single relaxation function $g_L(t)$ of \rf{GL}.  At the same time, the 
approximation  $S\rightarrow S_0$ is equivalent to replacing  $g(t)$ of \rf{f0} by  a simplified relaxation function with a single relaxation time, $g(t) \rightarrow  g_0(t)$, where
\beq{g0}
g_0(t) = \frac{1}{\tau_0}\, e^{-t/\tau_0} . 
\eeq
Therefore, rather than comparing the frequency domain functions, $S$, $S_0$ and $S_L$, one can equally well look at the time dependent relaxation functions $g(t)$ of \rf{f0}, $g_L(t)$ of \rf{GL}, and $g_0(t)$ of \rf{g0}.  These are compared in  Figure \ref{fx2}.

The  relaxation functions $g$, $g_0$  and $g_L$   share the common property that their integral is unity, 
\beq{in}
\int\limits_0^\infty g(t)dt =  \int\limits_0^\infty g_0(t)dt =\int\limits_0^\infty g_L(t)dt = 1 ,
\eeq
but the three functions  have distinct  first moments,  
\beq{in2}
\int\limits_0^\infty \, t g(t)dt =  \frac{\pi^2}{10}\, \tau_0 ,
\qquad \int\limits_0^\infty \, tg_0(t)dt =  \tau_0 ,
\qquad \int\limits_0^\infty \, tg_L(t)dt =  \frac{\pi^2}{12}\, \tau_0 .
\eeq
The functions $g(t)$ and $g_L(t)$  have the same values at $t= 0$,
$g(0) = g_L(0) = 12/(\pi^2 \tau_0)$ while $ g_0(0) = 1/ \tau_0$, 
as evident from Figure \ref{fx2}. 
The identities \rf{in} and \rf{in2} are associated with the first two terms in the low frequency expansions of the Fourier transforms.  The exact transforms are $\widetilde{g}$, defined in \rf{ap1}, and $\widetilde{g}_0$ and $\widetilde{g}_L$, given by  
\beq{ft}
\widetilde{g}_0(\omega) = \frac{1}{1- i\omega \tau_0}, \qquad
\widetilde{g}_L(\omega) = \frac{1}{1- i\omega \tau_L}\, ,  
\eeq
and the low frequency expansions are
\beq{ft2}
\widetilde{g}(\omega) = 1+ i\omega \tau_0 \frac{\pi^2}{10} + \ldots \, , \qquad
\widetilde{g}_0(\omega) = 1+ i\omega \tau_0 + \ldots \, , \qquad
\widetilde{g}_L(\omega) = 1+ i\omega \tau_0 \frac{\pi^2}{12} + \ldots \, .  
\eeq
Hence, one reason that $g_0(t)$ and $S_0 (\omega)$ are much better approximations than $g_L(t)$ and $S_L(\omega)$ to the precise quantities $g(t)$ and $S(\omega)$, respectively,  is that the number $\pi^2 /10 = 0.987$ is closer to unity than $\pi^2 /12 = 0.822$. 

These results imply that the Lagnese and Lions model could be improved by replacing the 
 term $(h/I)K \Theta = (12/h^2)K \Theta$ in \rf{ch5} with $(10/h^2)K \Theta$.  The resulting simplified single plate equation, analogous to Norris's equation \rf{n1} is then 
 \beq{LLN2}
(1+\eps_\nu )\, D\Delta^2 w - \eps_\nu\, D\, g_0*\Delta^2 w
 + \rho h w_{tt} = 0 .  
\eeq 
Note the relaxation function is now $g_0$, rather than $g_L$ as in eq. \rf{LLN}. The use of $g_L$ produces the inaccuracies shown in Figures \ref{fx1} and \ref{fx2}, while the same Figures illustrate that $g_0$ is a far better approximation to $g(t)$.  

The derivation of eq. \rf{LLN2} as an improvement on \rf{LLN} indicates that the 
source of the discrepancy in the Lagnese and Lions model can be partly ascribed to the term $(h/I)K \Theta $ in eq. \rf{llzb}.  
As noted before, the term in question 
originates in the   evaluation of $(K/I)\mu $ where the first moment of $ - \theta_{zz}$ is defined in \rf{th2} and evaluated, according to the linear approximation for $\theta$ assumed by Lagnese and Lions, in \rf{lin}.  Equation  \rf{lin} is incorrect.   The correct form follows from Norris and Photiadis \cite{Norris04}, and is most readily expressed in the frequency domain, as
\beq{lin2}
\widetilde{\theta} (x,y,z,\omega )  = \frac{1}{1- \widetilde{g}(\omega) }\, \big( z- \frac{\sin \gamma z}{\gamma \cos \gamma \frac{h}{2} }\big) \, \widetilde{\Theta}(x,y,\omega ) , 
\eeq
where $\gamma$ and $\widetilde{g}(\omega)$ are defined in eqs. \rf{defk} and \rf{ap1}, respectively. Thus, 
eqs. \rf{th2} and \rf{lin2} imply 
\beq{lin3}
\widetilde{\mu}  =  \big[\frac{1 -   \big(\gamma \frac{h}{2}\big)^{-1} \tan \gamma \frac{h}{2} }{1- \widetilde{g}(\omega) }\big] \, h \widetilde{\Theta} , 
\eeq
and expanding, 
\beq{lin4}
\widetilde{\mu}  = \big[ \frac{5}{6}  - \frac{(\gamma h)^2}{1008} +  \mbox{O}  \big(\gamma^4 h^4 \big) \big] \, h \widetilde{\Theta} . 
\eeq
This suggests that a better approximation than that of Lagnese and Lions is to take the leading order term, 
\beq{lin5}
\mu = \frac{5}{6}\, h\Theta, 
\eeq
which in turn implies the modified governing equation \rf{LLN2}. 

What are the relative benefits of the four different models?  The results presented indicate that for the infinite plate, one might just as well use the Norris model, since it is almost the simplest (only Simmond's is simpler) yet it provides the same accuracy as that of Chadwick.  The equivalence rests on the fact that for any practical plate of thickness $h$, one must have $h \gg l$, the mean free path of the thermal phonons.  It might be argued that the other models are more appropriate for finite systems, i.e. plates with edges subject to a variety of boundary conditions.  We have not discussed the edge conditions for each model, as this is a large topic in itself \cite{Lagnese88,Norris05b}.  Suffice to say that for each model one can define appropriate conditions that supplement the governing equations and define a closed problem for a given practical situation.  

In order to examine the utility of the models for a finite plate, Appendix D discusses the solution for the case of a beam with fixed ends and zero heat flux at the ends within the context of the Lagnese and Lions model.  The results there indicate that the additional equation \rf{llzb} for $\Theta $ only effects the solution in a boundary layer at the edges.  The Norris and Simmonds models do not provide a boundary layer solution, but have solutions analogous to the classical Kirchhoff theory but with modified (complex valued) flexural wavenumber.  Once could argue that the Lagnese and Lions model is useful in that it provides a boundary layer, even though the layer, being exponentially narrow, has insignificant effect on the value of both the modal frequencies and damping of the oscillating modes (see \cite{Norris05b} for further details).  The presence of a boundary layer at the edges is expected, and while the Lagnese and Lions model predicts one,  only the Chadwick model is capable of providing the proper description of the boundary layer.  Thus, the width of the boundary layer of the Lagnese and Lions model is O$(\delta^{1/2}/\kappa )$ where $\delta$ is defined in eq. \rf{ss} and $\kappa$ is the classical flexural wavenumber.  This boundary layer width could be less than, on the order of, or larger than the plate thickness, depending on the material parameters.  One can  envisage circumstances in which the boundary layer width is less than $h$, and consequently will have $z-$ dependence.  This is not contained within the  Lagnese and Lions model and can only be captured by the Chadwick model.  This fundamental difficulty arises with all beam and shell theories, classical or higher-order:  they do not accurately describe edge boundary layers that vary over a distance O$(h)$.  However, regardless of the precise nature of the temperature field near the boundary, it will  have essentially no effect upon the modal properties of the plate (frequency, damping, mode shape, etc.).

\section{Conclusion}

We have compared   four distinct models for flexural motion of thermoelastic thin plates, with specific attention to the leading order effects on wave damping.  Of the four, the Chadwick theory is expected to be the most reliable as it retains the 3D nature of the thermal field.  While the Lagnese and Lions has the same overall behavior as Chadwick's in terms of frequency, there is a distinct  quantitative difference.  This is attributable to the approximation of the first moment of the temperature, which is inaccurately accounted for by the Lagnese and Lions model. The relatively simple single equation of Simmonds predicts the correct low frequency behavior but is incorrect for medium and high frequencies.  The reason is that the Simmonds equations approximates the thermoelastic effect as a simple viscous damping term which does not capture the full frequency content. Comparisons of the Norris model with that of Chadwick indicates that the former adequately predicts the full frequency behavior.  The  single  equation of Norris provides an accurate approximation because the the dominant variation in the temperature field is in the through thickness direction, which permits it to be replaced by an equivalent viscoelastic term.   The Norris model retains the simplicity of a single uncoupled equation for plate deflection, but provides a more accurate representation of the thermoelastic ``hidden variables'' than Simmonds'.  The Lagnese and Lions model misses the crucial point that the  dependence of $\theta$ on the in-plane coordinates $x$ and $y$ is  secondary to its  dependence upon $z$.   This permits considerable simplification with the result that a separate equation for temperature is not necessary.  Instead, the temperature is incorporated into the classical plate equation as an effective viscoelastic relaxation \cite{Norris05b}.  
Of the four models, the simpler single equation of Norris captures  the thermal mechanics of the more general Chadwick theory, and is to be preferred to the other models considered.    

\appendix{A}
\section{Appendix A: A correction to the Lagnese and Lions equations}

In this Appendix the  correction to eqs. \rf{ch4} is obtained by the same method used in the original derivation of  Lagnese and Lions \cite{Lagnese88}. 
 The cause of the Lagnese and Lions error is that the  energy density $U$ (different notation) is defined in \cite{Lagnese88} by $U =\frac12  \bosy{\sigma}\cdot  \bosy{\varepsilon}$.  The correct energy density for the conjugate independent variables strain and temperature is the Helmholtz free energy $F$, see e.g. \cite{Norris04}, where
 $F = \frac12 (\bosy{\sigma}\cdot  \bosy{\varepsilon} -s \theta )$.  Using equations \rf{app0} and \rf{app4}, the total strain energy becomes  
\beq{app9}
{\cal F} = \frac{1}{2}\int\limits_\Omega \bigg\{
D \big[ (\Delta w)^2  - 2(1-\nu)\big(w_{11}w_{22}-w_{12}^2\big)
+ 2(1+\nu)\alpha \theta \Delta w  \big] - \frac{C_{}}{T} \theta^2\bigg\} dx dy\, . 
\eeq
The analogous equation (6.7) in  \cite{Lagnese88}  omits the final term and the coefficient of the $(1+\nu)\alpha \theta \Delta w$ term is unity rather than $2$, and hence the error in \rf{ch4}. 

\appendix{B}
\section{Appendix B: }

The functions defined by \rf{x2} may be found recursively as follows,
\beq{qa1}
u^{(n)}(s) = \sum\limits_{j=0}^n\, a^n_j\, \frac{s^{2n-2j+1} }{2n-2j+1}, 
\eeq
with 
\beq{qa2}
a^n_j = \frac{a^{n-1}_j}{(2n-2j)(2n-2j-1)} , \, \,  n > j \ge 0; \qquad 
a^n_n = - \sum\limits_{j=0}^{n-1}\, a^n_j , \, \,  n > 0; \qquad 
a^0_0=1.
\eeq

The values  $u^{(n)}(1)$ in eq. \rf{s61} can be found using these recursion relations.    We now demonstrate another, more constructive, way to obtain these coefficients.  The series expansion \rf{s61} may be obtained from the Norris solution for $\Theta$ written in the frequency domain as
\beq{bb2}
\widetilde{\Theta}(x,y,\omega)  = \frac{\alpha T E}{C(1-\nu )}\, \big[ 1- \widetilde{g}(\omega)\big]\, \Delta \widetilde{w}, 
\eeq
where $\widetilde{g}(\omega)$ is given in \rf{ap1}.  The latter may be expanded in a Taylor series as 
\beq{bb3}
\widetilde{g}(\omega) = 
\sum\limits_{m=0}^\infty  \,S_m \, (i\omega)^m, \qquad \mbox{where}\quad
S_m = \sum\limits_{n=0}^\infty  \, d_n \tau_n^m\, . 
\eeq
The infinite series $S_m$ may be evaluated using \rf{mod1}, and known results \cite{Mathematica}, as
\beq{bb4}
S_m = \tau_0^m\, 48 \, \pi^{2m}  \big( 2^{2m+4} - 1 \big)\, \frac{|B_{2m+4}|}{ (2m+4)!}\, , 
\eeq
where $B_m$ are the Bernoulli numbers \cite{Gradshteyn}.  Eliminating $\tau_0$ in favor of $\tau_S$ of \rf{x31} gives
\beq{gf}
S_m = 3 \, \tau_S^m\,  2^{2m+4}  \big( 2^{2m+4} - 1 \big)\, \frac{|B_{2m+4}|}{ (2m+4)!}\, , 
\eeq
and hence 
\beq{bb5}
\widetilde{g}(\omega) = 1 + \frac25 \, i\omega \tau_S  + \frac{17}{105}\, (i\omega \tau_S)^2 
+ \frac{62}{945}\, (i\omega \tau_S)^3  + \ldots . 
\eeq
Equation \rf{s61} then follows from \rf{bb2}, \rf{bb5} and the connection $-i\omega \rightarrow \partial /\partial t$.  Finally,  convergence properties of the series $\rf{bb2}_1$ follow from the behavior of the Bernoulli numbers for large index.  Omitting the details, it may be shown that the series is unconditionally convergent for $\omega \tau_0 < e$.

\appendix{C}
\section{Appendix C: Alternative justification of the $F_0$ approximation}
The justification for the approximation of $F_0$ (eqs. \rf{funf1} and \rf{f1}) is discussed further.  
If the reader is uncomfortable with the argument based on  the requirement that the phonon mean free path must be much shorter than the plate thickness, an alternative reason is given here.  We focus directly on the damping defined in  \rf{q} as given by eq. \rf{disp1}.  Thus, 
\beq{b1}
Q^{-1}  = 
  -  \eps_\nu   \pi^2\, \omega \tau_0\,   \mbox{Re}\, F_0\, , 
\eeq
where $F_0$ is given by \rf{funf1}. 

The Lagnese and Lions model is considered first.  It follows from \rf{b1} that
\beq{b2}
Q^{-1}  = 
  \eps_\nu   \frac{\omega \tau_* }{1+ \omega^2 \tau_*^2}, 
  \quad \mbox{Lagnese and Lions}, 
\eeq
where
\beq{b3}
\tau_*  =  \frac{\pi^2\, \tau_0}{12 + (\kappa h)^2}. 
\eeq
The thin plate model, whether thermoelastic or classical, is premised on the long wavelength assumption that 
\beq{b4}
\kappa h \ll 1.  
\eeq
Therefore,  
\beq{b5}
\tau_* =    \frac{\pi^2}{12 }\, \tau_0 + \mbox{O}[ (\kappa h)^2] \, ,  
\eeq
and it is entirely consistent with the thin plate theory (Kirchhoff approximation) to take 
$\tau_* =   \tau_0 \pi^2/12$, i.e., $\tau_* =\tau_L$. 

For the Chadwick model, eqs. \rf{q}, \rf{disp1} and \rf{ap1} imply 
\beq{b7}
Q^{-1}  = 
  \eps_\nu   \sum\limits_{n=0}^\infty  \,\frac{d_n\omega \tau_{n*} }{1+ \omega^2 \tau_{n*}^2}, 
  \quad \mbox{Chadwick}, 
\eeq
with
\beq{b8}
\tau_{n*}  =  \frac{ \tau_n}{1 + (\frac{\kappa h}{\pi})^2 \frac{\tau_n}{\tau_0}}. 
\eeq
Invoking the necessary condition \rf{b4}, we have  
\beq{b9}
\tau_{n*} =    \tau_n + \mbox{O} (\kappa^2 h^2) \, ,  
\eeq
and consequently, to the same order of accuracy, i.e. ignoring O$(\kappa^2 h^2)$, 
\beq{b10}
Q^{-1}  = 
  \eps_\nu   \sum\limits_{n=0}^\infty  \,\frac{d_n\omega \tau_{n} }{1+ \omega^2 \tau_{n}^2}, 
  \quad \mbox{Chadwick}. 
\eeq

\appendix{D}
\section{Appendix D: Solutions for a finite plate or beam}

We consider a plate of finite extent in the $x-$direction, $-L\le x \le L,\, -\infty < y < \infty$, with motion independent of $y$.  The ends are assumed, for simplicity, as fixed, with zero heat flux.  The problem is completely analogous to that of a beam, which is simply the limit of a plate of vanishing width in the $y-$direction \cite{norris03a}.  The beam solution is obtained   by setting $\nu \rightarrow 0$ but for consistency we will stay with the plate problem here.

For simplicity, we only seek solutions that are symmetric in $x$, i.e. of the form 
\beq{ec2}
 w(x,t) = \sum\limits_{i=1}^3 A_i\, \cos (k_i x - \omega t),
\qquad 
 \Theta (x,t)  = \sum\limits_{i=1}^3 \rho_i A_i\, \cos (k_i x - \omega t),
\eeq
where $\pm k_i$, $i=1,2,3$ are the three independent roots of \rf{disp} and $\rho_i$ follow from 
 \rf{llz}  as
\beq{ec4}
\rho_i  = \frac{k_i^4 - \kappa^4}{k_i^2 \alpha (1+\nu) }
 = \frac{\eps_\nu \gamma^2 k_i^2}{ \alpha (1+\nu) (k_i^2 - \gamma^2 +\frac{12}{h^2}) }\, . 
\eeq 
The end conditions 
\beq{ec1}
w(\pm L,t) = w_x (\pm L,t) = \theta_x (\pm L,t ) = 0\, , 
\eeq
imply 
\beqa{ec51}
\left[ 
\ba{ccc}
\cos k_1 L & \cos k_2 L & \cos k_3 L \\
& & \\
k_1\sin k_1 L & k_2\sin k_2 L & k_3\sin k_3 L \\
& & \\
\rho_1 k_1\sin k_1 L & \rho_2 k_2\sin k_2 L & \rho_3 k_3\sin k_3 L
\ea\right]
\left[ 
\ba{c}
A_1 \\ \\ A_2 \\ \\ A_3 \ea\right]
= 0\, , 
\eeqa
and in order that we have a non-trivial solution, $(\omega ,\, k)$ must satisfy  the dispersion relation 
\beqa{ec5}
\left| 
\ba{ccc}
\cos k_1 L & \cos k_2 L & \cos k_3 L \\
& & \\
k_1\sin k_1 L & k_2\sin k_2 L & k_3\sin k_3 L \\
& & \\
\rho_1 k_1\sin k_1 L & \rho_2 k_2\sin k_2 L & \rho_3 k_3\sin k_3 L
\ea\right| = 0\, . 
\eeqa

Let $k_1$ and $k_2$ correspond to the roots which are close to (within order $\eps_\nu$) the elastic roots $\pm \kappa$ and $\pm i \kappa$ (i.e. the roots of $k^4-\kappa^4=0$).  
The wavenumber $k_3$ is, to leading order, given by\footnote{This follows, for instance, by equating the two expressions for $\rho_3$ in \rf{ec4}.  Note that it does not assume any scaling between $\gamma$ and $1/h$, since  $\gamma h = $O$(\kappa h /\delta^{1/2})$, where both $\kappa h$ and $\delta$ are small quantities.} 
\beq{ec11}
k_3^2 \approx \gamma^2 - \frac{12}{h^2}\, . 
\eeq
It follows that 
\beq{ec61}
\left| \frac{\rho_1}{ \rho_3}\right|,\, \left| \frac{\rho_2}{ \rho_3}\right| \approx
\left| \frac{\eps_\nu \kappa^2 \gamma^2}{ k_3^2(\gamma^2 - \frac{12}{h^2} )} \right|  = 
\mbox{O}\big(\eps_\nu \delta \big), 
\eeq
where $\delta \ll 1$ is defined in \rf{ss}. Thus, 
\beq{ec6}
|\rho_1|, \, |\rho_2| \ll |\rho_3|,
\eeq
which enables approximation of eq. \rf{ec5} as 
\beqa{ec52}
\left| 
\ba{cc}
\cos k_1 L & \cos k_2 L  \\
&  \\
k_1\sin k_1 L & k_2\sin k_2 L 
\ea\right| = 0\, . 
\eeqa
To be more precise, $k_1, \, k_2$ are roots of $k^4-{\kappa '}^{4}=0$ where the complex valued wavenumber $\kappa '$ follows from \rf{dp}, 
\beq{dpp}
{\kappa '}^{4}  = \kappa^4 \big[ 
1 -  \eps_\nu\,  f ( \gamma h)\big]  ,  
\eeq
and $f$ is defined in \rf{f1}.  Thus, to leading order in the thermoelasticity the dispersion relation is, from \rf{ec52}, 
\beq{ec7}
\tan \kappa  'L  + \tanh \kappa ' L =  0\, . 
\eeq

It is important to note that this dispersion relation has the same form as that of the elastic plate or beam  without thermoelasticity.  The latter is obtained by the reduction $\kappa '\rightarrow \kappa$.  Thermoelasticity has the effect of modifying the wavenumber in the same way that it is altered in the infinite system,   and the end conditions do not introduce any new effects.  In fact, it can be easily checked that the dispersion relation \rf{ec7} is also obtained if one uses the Norris equation, which does not include temperature explicitly (and therefore does not require an end condition associated with temperature). 

The above result can be explained in terms of  boundary layers at the two ends.  In order to see this from the full solution for the Lagnese \& Lions model, we note that   $\rho_1$ and $\rho_2$ may be approximated using \rf{smd}, and combined with  \rf{ec51}  the leading order solution can be shown to be 
\beqa{ec8}
w&\approx & A_0\bigg( 
\sinh \kappa L \cos \kappa x + \sin \kappa L \cosh \kappa x + \frac{\eps_\nu 2\kappa^3 \gamma^2}{k_3^3(\gamma^2 - \frac{12}{h^2} )} \sin \kappa L \sinh \kappa L \frac{ \cos k_3 x}{\sin k_3 L} \bigg), 
\nonumber \\ && 
\\
\Theta &\approx & 
\frac{\eps_\nu 2 \kappa^2 \gamma^2 A_0}{\alpha (1+\nu) (\gamma^2 - \frac{12}{h^2} )} 
\bigg( -
\sinh \kappa L \cos \kappa x + \sin \kappa L \cosh \kappa x + \frac{2\kappa}{k_3} \sin \kappa L \sinh \kappa L \frac{ \cos k_3 x}{\sin k_3 L} \bigg). 
\nonumber
\eeqa
With no loss in generality, choose $k_3$ as the root with positive imaginary part.  Then $|e^{ik_3L}| \ll |e^{-ik_3L}|$, and consequently, for positions not near the center of the plate, we have 
\beq{ec12}
\Theta \approx  
\frac{\eps_\nu 2 \kappa^2 \gamma^2 A_0}{\alpha (1+\nu) (\gamma^2 - \frac{12}{h^2} )} 
\bigg( -
\sinh \kappa L \cos \kappa x + \sin \kappa L \cosh \kappa x -i\frac{\kappa}{k_3} \sin \kappa L \sinh \kappa L 
\, e^{ik_3(L-|x|)} \bigg), 
\eeq
with a similar equation for $w$. 
While \rf{ec12}  is not valid at the center $(x=0)$ it provides a  faithful leading order approximation near the ends.  The decay is far more rapid than that of the evanescent flexural wave, since Im$(k_3/\kappa )\gg 1$ (see eqs. \rf{ss} and \rf{smd}). This  indicates that the contribution of the  complex wavenumber $k_3$ is an exponentially decaying boundary layer.  The  fact that $|k_3|h$ is on the order of or larger than unity raises doubts about the utility of the model, particularly near boundaries. 

\bigskip
\noindent
{\bf \large Acknowledgment}: Comments from Prof. James G. Simmonds are appreciated. 


\end{document}